\documentclass[twocolumn, trackchanges]{aastex63}

\usepackage{graphicx, amsmath}
\usepackage[caption=false]{subfig}
\usepackage{hyperref}
\usepackage{cleveref}
\usepackage{xspace}
\usepackage{rotating}

\usepackage{comment}

\newcommand{\lya}{Ly$\alpha$\ } 

\newcommand{\lyaa}{Ly$\alpha$} 

\newcommand{\OII}{[\ion{O}{2}]\xspace}

\newcommand{\escma}{erg~s$^{-1}$~cm$^{-2}$~\AA$^{-1}$}

\graphicspath{{./}{figures/}}

\begin{document}
\title{The HETDEX Survey: Probing neutral hydrogen in the circumgalactic medium of $\sim$88,000 Lyman Alpha Emitters}

\author[0009-0003-1893-9526]{Mahan Mirza Khanlari}
\affiliation{Department of Astronomy, The University of Texas at Austin, 2515 Speedway Boulevard, Stop C1400, Austin, TX 78712, USA}

\author[0000-0002-8433-8185]{Karl Gebhardt}
\affiliation{Department of Astronomy, The University of Texas at Austin, 2515 Speedway Boulevard, Stop C1400, Austin, TX 78712, USA}

\author[0000-0002-4974-1243]{Laurel H. Weiss}
\affiliation{Department of Astronomy, The University of Texas at Austin, 2515 Speedway Boulevard, Stop C1400, Austin, TX 78712, USA}

\author[0000-0002-8925-9769]{Dustin Davis}
\affiliation{Department of Astronomy, The University of Texas at Austin, 2515 Speedway Boulevard, Stop C1400, Austin, TX 78712, USA}

\author[0000-0002-2307-0146]{Erin Mentuch Cooper}
\affiliation{Department of Astronomy, The University of Texas at Austin, 2515 Speedway Boulevard, Stop C1400, Austin, TX 78712, USA}
\affiliation{McDonald Observatory, The University of Texas at Austin, 2515 Speedway Boulevard, Stop C1402, Austin, TX 78712, USA}

\author[0000-0001-7066-1240]{Mahdi Qezlou}
\affiliation{Department of Astronomy, The University of Texas at Austin, 2515 Speedway Boulevard, Stop C1400, Austin, TX 78712, USA}

\author[0000-0002-6907-8370]{Maja {Lujan Niemeyer}}
\affiliation{Max-Planck-Institut f\"{u}r Astrophysik, Karl-Schwarzschild-Str. 1, 85741 Garching, Germany}

\author[0000-0002-1328-0211]{Robin Ciardullo} \affil{Department of Astronomy \& Astrophysics, The Pennsylvania State University, University Park, PA 16802, USA} \affil{Institute for Gravitation and the Cosmos, The Pennsylvania State University, University Park, PA 16802, USA}

\author[0000-0001-7240-7449]{Donald P. Schneider}
\affiliation{Department of Astronomy \& Astrophysics, The Pennsylvania
State University, University Park, PA 16802, USA}
\affiliation{Institute for Gravitation and the Cosmos, The Pennsylvania State University, University Park, PA 16802, USA}

\author[0000-0003-3823-8279]{Shiro Mukae}
\affiliation{Department of Astronomy, The University of Texas at Austin, 2515 Speedway Boulevard, Stop C1400, Austin, TX 78712, USA}

\author[0000-0001-5561-2010]{Chenxu Liu}
\affiliation{South-Western Institute for Astronomy Research, Yunnan University, Kunming, Yunnan, 650500, People’s Republic of China}
\affiliation{Department of Astronomy, The University of Texas at Austin, 2515 Speedway Boulevard, Stop C1400, Austin, TX 78712, USA}

\author[0000-0003-2575-0652]{Daniel Farrow}
\affiliation{Centre of Excellence for Data Science,
Artificial Intelligence \& Modelling (DAIM),
University of Hull, Cottingham Road, Hull, HU6 7RX, UK}
\affiliation{E. A. Milne Centre for Astrophysics
University of Hull, Cottingham Road, Hull, HU6 7RX, UK}

\author[0000-0001-6717-7685]{Gary J. Hill} 
\affiliation{McDonald Observatory, The University of Texas at Austin, 2515 Speedway Boulevard, Stop C1402, Austin, TX 78712, USA}
\affiliation{Department of Astronomy, The University of Texas at Austin, 2515 Speedway Boulevard, Stop C1400, Austin, TX 78712, USA}

\author[0000-0003-2307-0629]{Gregory R. Zeimann}
\affiliation{Hobby Eberly Telescope, University of Texas, Austin, Austin, TX, 78712, USA}

\author[0000-0002-0417-1494]{Wolfram Kollatschny}
\affiliation{Institut f\"ur Astrophysik and Geophysik, Universit\"at G\"ottingen, Friedrich-Hund Platz 1, D-37077 G\"ottingen, Germany}

\begin{abstract}
We explore the neutral hydrogen (\ion{H}{1}) gas around $1.9 < z < 3.5$ Lyman Alpha Emitters (LAEs) from the Hobby-Eberly Telescope Dark Energy Experiment (HETDEX) using faint \lya absorption. This absorption is the result of \ion{H}{1} in the halo of the LAE scattering \lya photons from the integrated light of background galaxies along the line of sight. We stack millions of spectra from regions around $\sim$ 88,000 LAEs, in order to focus on the physics of the gas at large radii. The extensive number of fiber spectra contributing to the stacks ensures significant signal-to-noise ratio (S/N) to detect the faint \lya absorption which would otherwise be buried within the noise. We detect absorption out to a projected $\sim$  350 kpc around an average LAE at \textit{z} $\sim$ 2.5.  We use these results to create an empirical radial $W_{\lambda}$(\lyaa) profile around LAEs. Comparison with numerical simulations reveals a profile similar to the empirical one within this region. Compared to previous studies, the profile is similar but modestly higher. We also outline a simple physical picture motivated by the observed trends in the data. We plan to quantify this radial profile as a function of redshift, local density, and \lya luminosity to  explore the relationship between LAE environments and \ion{H}{1} distribution.
\end{abstract}

\keywords{cosmology: observations -- galaxies: structure -- galaxies: evolution -- galaxies: high redshift}
 
\section{\textbf{Introduction}}\label{sec:intro}

Understanding the distribution and dynamics of diffuse neutral hydrogen (\ion{H}{1}) gas within the circumgalactic medium (CGM) and the intergalactic medium (IGM) is fundamental to cosmology and the study of galaxy evolution. Hydrogen, which constitutes the main portion of the universe's baryonic mass, plays a pivotal role in the cosmic life cycle by fueling star formation and governing the interplay between galaxies and large-scale structures \citep{white_core_1978}. The CGM acts as an interface between galaxies and the IGM, encapsulating the complex processes of accretion, feedback, and structure evolution. The CGM therefore is a unique tool for studying the cycle of baryonic matter (See \citealt{tumlinson_circumgalactic_2017} and \citealt{Faucher_CGM_rev} for a review).

The Hubble Space Telescope has produced detailed, high-resolution observations of the CGM in the local universe \citep{werk_cos-halos_2013,Tumlinson_2011}. These studies, which use absorption lines in the spectra of background quasars, shed light on the ionization fraction and chemical makeup of the gas surrounding nearby galaxies. Careful examination of these interactions is essential for understanding the gas cycles within galaxies \citep{nielsen_revealing_2023}. Local universe studies therefore play an important role in studying how the CGM and IGM contribute to galactic evolution and the cycle of baryonic matter.

The \lya forest \citep{gunn_density_1965} is also crucial for exploring the IGM and the CGM of galaxies that intersect quasar (or star-forming galaxy) sightlines \citep{rauch_lyman_1998}. Such studies reveal the gas distribution and condition in the CGM (e.g. \citealt{Rudie_CGM_2012,Steidel_2010,Muzahid_2021}). However, the effectiveness of the \lya forest as a probe for CGM absorbers is limited by the need for specific alignments with quasars, narrowing the scope of environmental studies around galaxies.  As a result, this method struggles to provide a complete statistical view of CGM absorbers, especially at higher redshifts where only small patches of large-scale structure are illuminated by background quasars.

In response to these challenges-- where the rarity of bright quasars pushes us to rely on fainter galaxies as our background source -- we turn to stacking spectra as a powerful solution.  \citealt{Weiss_2025} has shown that this background light is indeed detectable in HETDEX. Background galaxies are intrinsically fainter relative to quasars, which makes it difficult to detect subtle features in individual spectra. Stacking helps by boosting the signal-to-noise ratio (S/N), which makes it possible to detect faint spectral features that would otherwise be lost in the noise \citep{shapley_detailed_2003, davis_hetdex_2023-1}. Both \lya emission and absorption have been probed around high redshift star-forming galaxies through spectroscopic stacking. In emission, stacked spectra have detected \lya extending out to 80 kpc \citep{Steidel_2010}, 100-200 kpc \citep{Zhang_2024}, and even as far as 320 kpc \citep{niemeyer_surface_2022}. In absorption, studies employing both background quasars and star-forming galaxies have mapped the \lya\ absorption profiles in the CGM \citep{Chen_2020,Steidel_2011,Muzahid_2021,Lofthouse_2020,Banerjee_2025}. Notably, at large distances from the center of the star-forming galaxies, the extended \lya emission from heavy star formation diminishes, and \ion{H}{1} can still be traced via \lya absorption from background  light \citep{Matthee_2024}.

The catalog of over 800,000 LAEs in the Hobby-Eberly Telescope Dark Energy Experiment \citep[HETDEX;][]{gebhardt_hobby-eberly_2021} offers a particular advantage for stacking, given the sheer number of spectra. Beyond the large number of spectra, HETDEX’s wide-area provides on-sky fiber coverage around each LAE. This coverage allows for extensive investigation into the CGM of these LAEs. Such a large sample offers the statistical robustness required to analyze \ion{H}{1} gas across different environments and redshifts, presenting a perspective which would otherwise be unachievable in quasar sightline studies. As a result, by leveraging the large number of LAEs in the HETDEX dataset (and the even larger number of fibers around them) we can explore variations in \ion{H}{1} density as a function of transverse distance from these galaxies. Additionally, we can investigate this trend for different environments within the large scale structure.

The paper is structured as follows: Section \ref{sec:data} briefly introduces HETDEX and the LAE dataset used in this paper. Section \ref{sec:stacking} describes the methodology for spectral stacking and the corrections prior to the stacking. Section \ref{sec:stacking_regions} defines the regions around the LAEs where we collect and stack fiber spectra. Section \ref{sec:ew_measure} describes the methodology of equivalent width measurement. Section \ref{sec:results} compares the observed \ion{H}{1} radial profile to numerical simulations and previous studies. Section \ref{sec:discussion} discusses a potential physical picture of the absorption and the observed trends in the dataset, the implications of our findings, and the future directions. And Section \ref{sec:summary} provides a summary of the key results. Throughout this study, we adhere to the Planck 2018 cosmology \citep{planck_collaboration_i_planck_2020}, characterized by $\Omega_{\text{m}}$= 0.31 and $H_0$ = 67.7 $\mathrm{km~s^{-1}~Mpc^{-1}}$.

\begin{figure*}
    \centering
    \includegraphics[height=5.8cm]{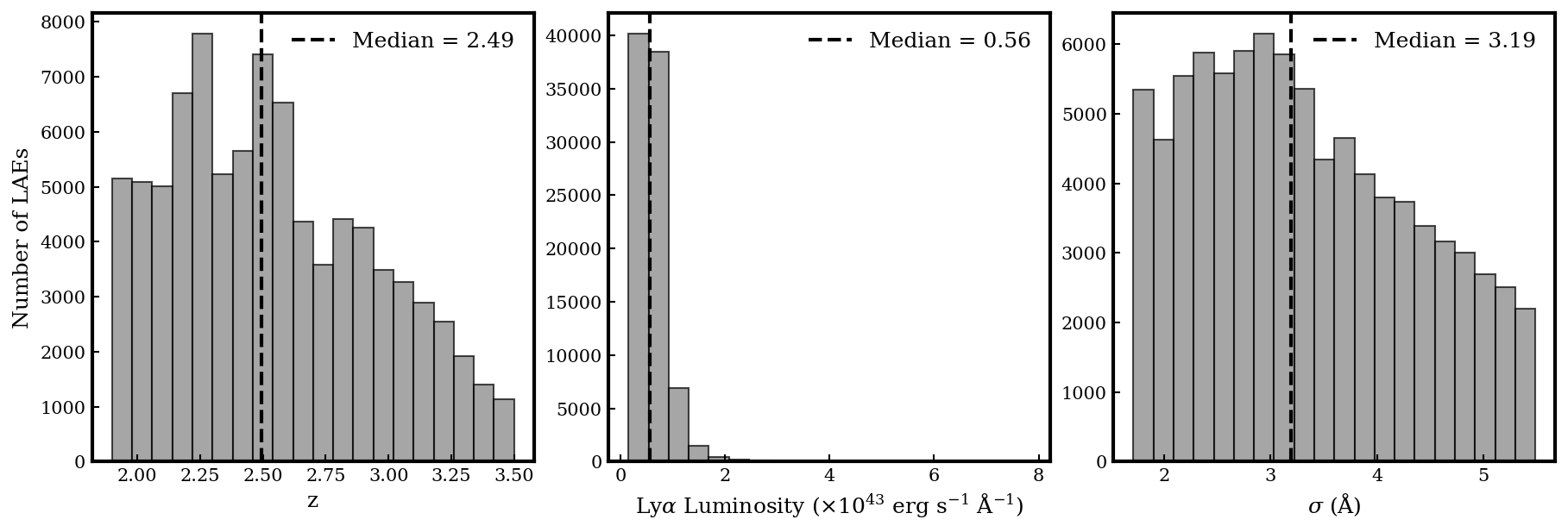}
    \caption{Histograms displaying the key properties of the HETDEX LAEs. The panels (from left to right) show the redshift, \lya luminosity, and line width distributions, with dashed lines indicating each parameter’s median value.}
    \label{fig:hists}
\end{figure*}

\section{\textbf{Data}}
\label{sec:data}

HETDEX \citep{gebhardt_hobby-eberly_2021} is an untargeted wide-field spectroscopic survey that utilizes the upgraded Hobby–Eberly Telescope (HET; \citealt{ramsey_early_1998} \citealt{hill_completion_2018}). The survey collected data from 2017 through 2024 using the Visible Integral-Field Replicable Unit Spectrograph \citep[VIRUS;][]{hill_hetdex_2021}, an instrument comprised of 78 Integral Field Units (IFUs) and 156 spectrographs, each containing 448 $1\farcs 5$-diameter fibers.  The survey principally covers two high-galactic latitude regions of the sky -- a ``Spring" field and ``Fall'' field -- with a total area of $\sim 540$~deg$^2$ at a fill factor of roughly 1:4.6  \citep{hill_virus_2018, hill_hetdex_2021}.  HETDEX data have also been acquired in a few science verification regions, such as the Cosmic Evolution Survey \citep[COSMOS;][]{scoville_cosmic_2007} and the Great Observatories Origins Deep Survey-North \citep[GOODS-N;][]{giavalisco_great_2004} fields.

The VIRUS IFUs are distributed in a grid-like pattern across the HET's focal plane, with each IFU separated from the others by $51\arcsec$, except in the center of the field, where other HET instruments are located. Each individual IFU captures a field of view of $51\arcsec \times 51\arcsec$ and feeds two spectrographs covering the wavelength range 3500 to 5500~\AA\ at resolving power between $750 < R < 950$.  Since these IFUs do not have lenslets, a three exposure dither pattern is used to fill in the gaps between fibers, and each exposure is typically $\sim 6.1$~min in length. A HETDEX observation therefore consists of 104,832 spectra (34,944 per exposure) distributed over a $\sim 21\arcmin$ science field-of-view. Spectra are obtained for all objects covered by the IFUs, including the \lya emitting galaxies in the redshift range $1.9 < z < 3.5$.

The processing of HETDEX frames is described in \cite{gebhardt_hobby-eberly_2021}.  Since HETDEX's observational strategy does not involve the pre-selection of targets, the experiment captures unbiased spectral data across its entire survey area. The processing pipeline for these spectra involves sky subtraction, astrometric and photometric calibration, object detection, and the integration of spectra from a contiguous set of fibers around each detected emission-line or continuum source into a single, PSF-weighted spectrum \citep{gebhardt_hobby-eberly_2021}. The detections are classified and a redshift is determined via the ELiXer software package \citep{davis_hetdex_2023}. 

Through this methodology, HETDEX has identified approximately 800,000 Lyman Alpha Emitters (LAEs) with a S/N $>5$ in HETDEX Data Release (HDR 4.0.0), encompassing data from January 2017 through to August 2023.  For this paper, we use fibers which extend over a large radius around each LAE to probe the surrounding diffuse \ion{H}{1} gas environment, and build upon the result of the negatively enhanced absorption troughs from the stacked LAE spectra presented in \cite{davis_hetdex_2023-1} and described in \cite{weiss_absorption_2024}.

By utilizing the fiber spectra, rather than the PSF-extracted spectra from the aperture, we are able to explore the \ion{H}{1} distribution around the LAEs in the dataset. This approach allows for a direct examination of the environment of these LAEs by analyzing the spectra of tens of millions of fibers associated with the $\sim$ 88,000 high-confidence LAE detections in the redshift range of $1.9< z <3.5$ . We limit the LAE sample to  S/N $>5$ , and we exclude LAEs with line widths $\sigma > 5.5$ \AA\ ($\sim 350$ km s$^{-1}$) to eliminate potential AGNs that are not otherwise identified in prior screenings or external catalogs (e.g., \citealt{liu_active_2022}). We further limit the sample to those that have been identified as an LAE via ELiXer and restrict the selection to observations with good seeing (i.e., point-spread function full width at half maximum, or FWHM $<1.5\arcsec$). We also apply an aperture correction threshold (quantified by ``apcor" $> 0.6$) to exclude LAEs located at the very edges of the IFUs. Figure \ref{fig:hists} shows the distribution of redshift, \lya line luminosity, and \lya line width ($\sigma$) of our LAE sample.

\section{\textbf{Spectral Stacking}}
\label{sec:stacking}
High S/N spectra are vital for detecting subtle physical signatures in high-redshift environments. Since achieving a high S/N for an individual spectrum is challenging, we employ the spectral stacking approach as outlined in \cite{davis_hetdex_2023-1} and \cite{weiss_absorption_2024}. A key difference, however, is while those studies stack spectra directly centered on the LAEs, we combine spectra from the area \textit{surrounding} the LAEs. We use the regions around tens of thousands of LAEs to significantly enhance our ability to detect \lya absorption. We explicitly exclude fiber spectra from the central emission line region to focus on the gas properties beyond the galaxy center. Further discussion regarding these regions will be described in Section \ref{sec:stacking_regions}.

\subsection{Pre-Stacking Spectral Corrections}
\label{sec:corrections}
Prior to stacking we apply several corrections to the HETDEX spectra. Firstly, individual HETDEX fiber spectra are sky subtracted according to the HETDEX sky subtraction procedure. Next, an improved sky background estimation using sky residuals is applied to each fiber. Finally, we select and remove fibers to mitigate the effects of atmospheric interference, foreground sources, and significantly bright background light.

\subsubsection{Sky Subtraction}
\label{sky_sub}
While the detailed sky subtraction methodology is presented in \citealt{gebhardt_hobby-eberly_2021}, we provide a brief overview due to its relevance to this work. The primary objective of sky subtraction is to eliminate excess light while preserving the signal from astronomical sources. This light includes atmospheric emission and instrumental effects such as thermal noise. In this paper, we use a full-field sky measurement across the HET's 21\arcmin\ diameter focal plane, utilizing all 34,944 fibers in each HETDEX observation (see \citealt{gebhardt_hobby-eberly_2021} and \citealt{niemeyer_surface_2022} for detailed explanations). We exclude fibers that contain source detections and/or noticeable continuum, originating from stars or bright galaxies. The threshold for exclusion is defined by flux levels exceeding $3\times$ the biweight scale \citep{beers_measures_1990} of the average of all fibers in a field. The remaining fibers are then averaged and subtracted from each fiber.

\subsubsection{Sky Residual Correction}
\label{sky_res}
To achieve a sky subtraction accuracy better than 1\%, we apply a per-fiber sky residual correction. This process is necessary for stacking measurements, where subtle residual biases that are negligible in individual spectra can become significant when stacking large numbers of spectra. The approach involves statistically compositing the spectrum of an ``empty'' fiber on a per-observation basis. This composite spectrum is then subtracted from the already full-field sky calibrated source spectra within the same observation to improve the removal of unwanted flux not originating from the target source. The definition of an ``empty'' fiber is based on the system's detection limits and observational conditions, recognizing that no fiber is completely free of photon sources along the line of sight.

For each observation, we load all of the full-field sky calibrated fiber spectra, excluding fibers that are flagged due to meteors or satellites, as well as possible instrumental issues, such as malfunctioning amplifiers or poor throughput. Fibers containing an excessive number of zeros or invalid values are also removed.

Next, we exclude fibers that contain continua or are excessively negative. This exclusion is based on the median flux density across defined wavelength ranges. Fibers are excluded if they have a median flux density $> -5 \times 10^{-19}$  or $<25 \times 10^{-19}$ \escma\ between 3500 \AA~ and 3860 \r{A}, or a median flux density that does not fall within $\pm 5 \times 10^{-19}$ \escma\ for the rest of the wavelength range (3860-5500 \r{A}). The remaining fibers, deemed free from significant continuum sources or extreme values, are then stacked to create the residual corrective stack. This corrective stack is then subtracted from each fiber spectrum in the observation, minimizing slight calibration errors and the influence of faint, extraneous light sources.

\begin{figure}
    \centering
    \includegraphics[height=5cm]{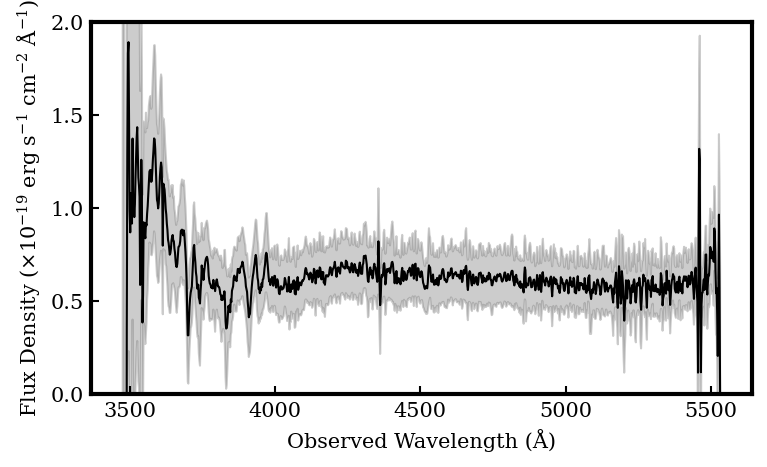}
    \caption{The biweight residual spectrum across approximately 1050 HETDEX observations. Roughly 70\% of fibers in each observation are used in computing the empty-fiber average, with the biweight scale shown in gray. Note the increased flux in the blue due to instrumental limitations. We reiterate that this residual level is about 1\% of the survey’s flux limit, which is irrelevant analyzing single spectra.}
    \label{fig:residuals}
\end{figure}

We also integrate a Hyper Suprime-Cam (HSC) \textit{g}-band interface that automates the selection of the optimal residual correction. Rather than manual specification, the pipeline cross-matches $\sim$ 60,000 HSC–\textit{g} continuum sources to HETDEX spectra, bins them by \textit{g}-band in 0.2 mag intervals, applies each candidate empty-fiber correction to individual spectra before stacking, and measures the \textit{g}-band magnitude of the resulting stacked spectra. The empty-fiber correction that minimizes the difference between the HETDEX computed \textit{g}-band and the HSC–\textit{g} photometry is selected automatically. Figure \ref{fig:residuals} shows the biweight average of the residuals produced by sky subtraction, derived from stacking more than 300,000 random empty fibers from about 1050 unique HETDEX observations. The errors, displayed in gray, reflect the biweight scale. The flux level of this residual sky spectrum is approximately 1\% of the HETDEX flux limits, which is only significant when stacking large numbers of spectra.

\subsubsection{Fiber Selection}
\label{subsec:fiber_sel}
Due to the faint nature of extended \lya absorption, we need to carefully select the fibers we use in our stacks around LAEs to mitigate the contribution of light from foreground and background sources. Foreground light includes any emission from sources at a lower redshift than the LAEs, such as stars or bright foreground galaxies. Additionally, while we aim to include diffuse illumination from populations of background galaxies, we must avoid direct line-of-sight emission from these galaxies. In practice, we identify and exclude discrete sources by setting thresholds for absolute flux density levels, as in Section \ref{sky_res}. We exclude fibers with a median measured flux density exceeding $5 \times 10^{-19}$ \escma\ across the entire spectral range of the full-field sky calibrated spectrum. Such fibers likely contain distinct, potentially resolvable sources and are therefore not included in the analysis.

\subsection{Stacking Methodology}\label{stack_methods}
\label{subsec:stack_methods}

Following these corrections, the spectra are converted from air to vacuum wavelengths \citep{greisen_representations_2006} and shifted to the rest frame of the associated LAE using HETDEX-derived redshifts \citep{davis_hetdex_2023,mentuch_cooper_hetdex_2023}. We then convert the observed flux densities to luminosity densities to account for cosmological dimming. This prevents a bias towards relatively lower redshift LAEs in a stack and allows for a consistent comparison of apparent luminosities across different redshifts (See \citealt{davis_hetdex_2023-1} for a detailed explanation).

Once we have obtained the corrected, rest frame spectra, we stack millions of fibers in regions surrounding the LAEs. The stacking process involves a linear interpolation of the contributing individual spectra onto a unified rest-frame wavelength grid. The extent of this grid is determined by the highest- and lowest-redshift objects in the sample, with the grid spacing adopted from the highest-redshift object. We then take a biweight statistic of the spectral points within each wavelength bin. This creates a single stacked spectrum. For a thorough explanation on this methodology, see \cite{davis_hetdex_2023} and  \cite{weiss_absorption_2024}.  This method, supported by the extensive HETDEX dataset, allows us to perform an averaged analysis across different orientations, environments, and geometrical configurations of LAEs. We will discuss how we define regions in which to stack in the next section. 

\section{\textbf{Target Regions Around the LAE}}
\label{sec:stacking_regions}

\begin{figure}
    \centering
    \includegraphics[height=8cm]{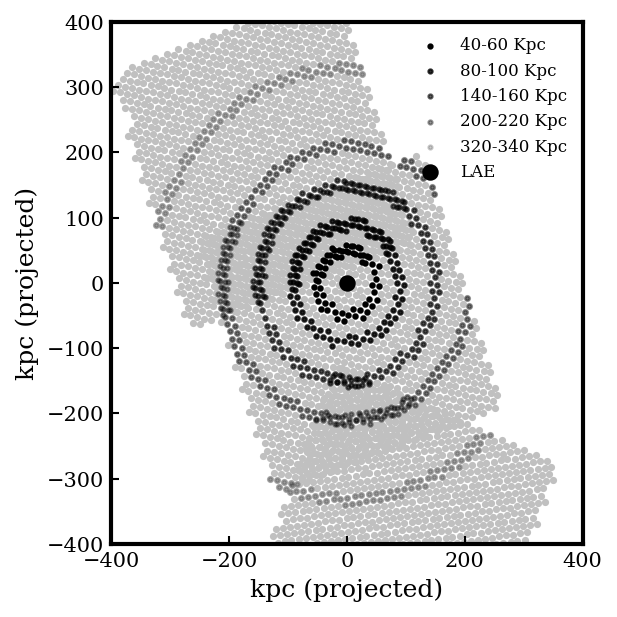}
    \caption{An example of how fiber spectra are extracted around a representative LAE\null. The squares of gray points indicate the IFUs, and in regions where IFUs overlap (mostly in science verification fields), fibers from neighboring IFUs are sometimes included. The IFUs are typically spaced $51\arcsec$ apart. The coordinates are converted to physical units based on the LAE’s redshift.}
    \label{fig:rings}
\end{figure}

Our primary analysis involves probing beyond the immediate vicinity of LAEs into larger regions of space. This section outlines the procedure and details of this approach. All the distances are physical, and we do not collect fibers from the inner 40 kpc of each LAE to avoid potentially capturing \lya emission that often surrounds $z \sim 2.5$ LAEs. This extent is also largely due to HETDEX's $1.5\arcsec$ diameter fibers, making probing the inner regions of LAEs difficult. Figure \ref{fig:rings} shows an example of collecting fibers around an LAE\null. As depicted, we collect and stack fibers within 20 kpc annuli, beginning at a transverse distance from the LAE, or $D_{tran}$, of 40 kpc. For display purposes, only a subset of these annuli are shown in the figure. These rings are not perfect circles since we lack complete sky coverage; occasionally, we collect fibers from an IFU other than the one containing the LAE.

For each annulus around an LAE, we compute the biweight average of all spectra from the vetted fibers (see Section \ref{subsec:fiber_sel}) that fall within the defined ring in the observed frame. The averaged spectrum from each annulus will then be shifted to the rest frame and used for the final biweight stacks of annuli around all LAEs in the sample.

\begin{figure}
    \centering
    \includegraphics[height=5.2cm]{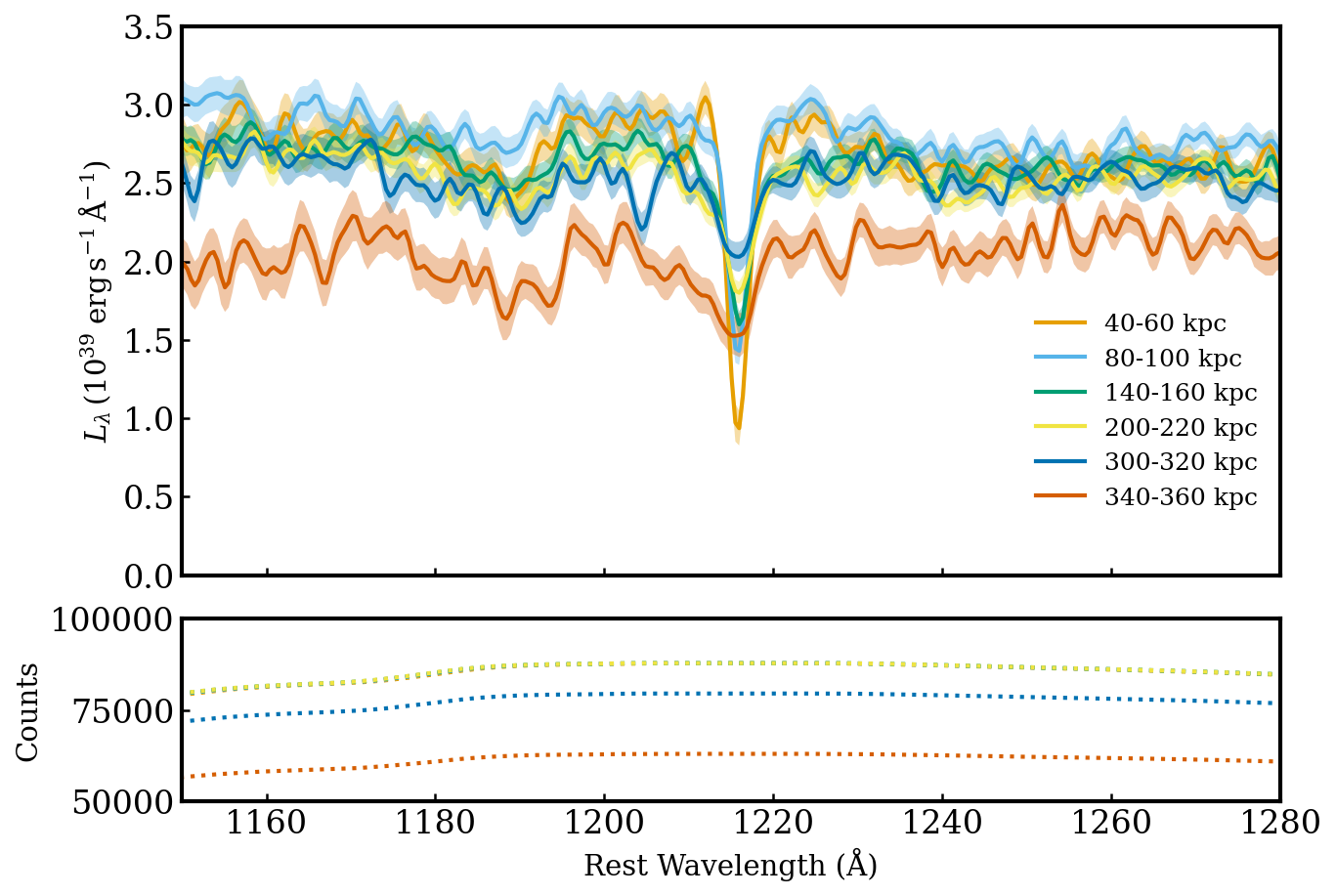}
    \caption{\textit{Top panel:} Stacked spectra derived from nearly 88,000 high-confidence LAEs, arranged by transverse distance ($D_{tran}$) from the galaxy center. Each spectrum is a biweight stack of $\sim$ 3 million individual spectra, and the final stack is smoothed with a 1-pixel Gaussian kernel. Lighter color bands are the $1\sigma$ error envelopes derived from 200 bootstrap resamplings of the galaxy sample.} The \lya absorption is apparent in each spectrum until the largest radial bin at $\sim350$\,kpc. The uncertainty on the flux values are consistent with the observed RMS in the continuum (about 5\% of the total continuum). \textit{Bottom panel:} The number of LAEs contributing to each wavelength bin.
    \label{fig:annuli}
\end{figure}

To determine the evolution of the \ion{H}{1} surrounding an average LAE at $z \sim 2.5$, we stack millions of spectra in annuli and examine the strength of the \lya absorption as a function of $D_{tran}$.  Figure \ref{fig:annuli} depicts stacked spectra of the environments of approximately 88,000 high-confidence LAEs at varying $D_{tran}$. Each spectrum is built from a biweight stack of $\sim$ 3 million individual spectra. We repeat the stacking with simple mean and median statistics and the change in the result is negligible. The stacks reveal clear absorption extending up to $D_{tran}$ of $\sim$ 350 kpc. We stress, however, that this is a statistical absorption signal against a statistical continuum. Lighter color bands show the $1\sigma$ error envelopes derived from 200 bootstrap resamplings of the galaxy sample.

The continuum detected in Figure \ref{fig:annuli} reflects the integrated light from galaxies behind the LAEs along the line of sight. The continuum level is relatively consistent with the \lya surface brightness profile reported by \citet{niemeyer_surface_2022}. Most of the flux of this background light comes from the local vicinity due to clustering. As we increase $D_{tran}$, the expectation is that the local density of background galaxies decreases, leading to a relatively lower continuum flux. The LAE, on average, will be centered on the local density peak, which results in a higher continuum flux at closer $D_{tran}$, though this increase is relatively small (see Figure \ref{fig:annuli}).

\section{\textbf{Equivalent Width Measurement}}
\label{sec:ew_measure}
After stacking the spectra in annuli from $\sim 88,000$ LAEs, our goal is to measure the rest-frame equivalent width ($W_{\lambda}$(\lyaa)) as a function of projected distance. Measuring $W_{\lambda}$(\lyaa) quantifies the strength of the absorption and allows for a direct comparison to simulations and previous studies. Throughout this work, unless otherwise stated, ``annuli'' refer to stacks of fibers at a given distance around the full LAE sample.

We normalize each stacked spectrum to its continuum, which we estimate using regions both blue-ward (1100–1150\,\AA) and red-ward (1300–1350\,\AA) of the \lya\ absorption line. We take the median flux within each region and interpolate between them to obtain an initial continuum estimate. We then fit a straight line to the interpolated values to create a smooth continuum model used for normalization. Next, to further boost S/N, we combine pairs of 20\,kpc bins by biweight averaging (e.g., merging the 40--60\,kpc and 60--80\,kpc annuli into a single 40--80\,kpc bin). Our final set of bins extends from 40\,kpc out to 360\,kpc. Merging 20\,kpc annuli to 40\,kpc bins retains the option to perform higher spatial resolution analysis if needed using the original 20\,kpc bins.

Attempts to fit the absorption with Gaussian or Voigt profiles are unable to adequately capture the line shapes, especially in larger annuli where the absorption profile becomes more irregular. Therefore, we adopt a non-parametric approach to measure  $W_{\lambda}$(\lyaa). For each measurement (e.g., each choice of continuum side bands and each integration range) from each stacked annulus, the equivalent width can be written as
\begin{equation}
    \mathrm{EW}_{ij} \;=\; 
    \sum_{k=1}^{M} \left(1 - \frac{F_{ijk}}{C_{ij}} \right) \Delta \lambda_{k},
\end{equation}

where $C_{ij}$ is the chosen continuum level for the $i$th annulus and $j$th continuum window definition, $F_{ijk}$ is the measured flux in each wavelength bin $k$, and $\Delta \lambda_{k}$ is the width of that bin. The total number of wavelength bins in the line region is $M$.

To quantify the uncertainties, we adopt a multi-step procedure that account for both statistical and systematic variations. First, for each $(i,j)$ pair, we propagate the flux errors (including continuum fitting uncertainties) to estimate a statistical variance, $\sigma_{\mathrm{stat},ij}^2$. Then, to account for the systematic uncertainty arising from the choice of continuum estimation regions and integration ranges, we vary the definitions of these regions and repeat the measurement across multiple combinations. Specifically, the blue continuum window begins from 1085--1095\,\AA\ and goes up to 1185--1195\,\AA\, maintaining 10\AA\ wide bins without overlap. Similarly, the red continuum window begins from 1225--1235\,\AA\ and goes up to 1325--1335\,\AA. The line integration region boundaries are also shifted, with the lower boundary moving from 1211\,\AA\ downward by 0.1\,\AA\ per step (ten steps total) and the upper boundary moving from 1219\,\AA\ upward by 0.1\,\AA\ per step. This procedure produces $N = n_{\mathrm{line}} \times n_{\mathrm{cont}} = 100$ unique measurements per annulus:
\begin{equation}
    \mathrm{EW}_{ij}, \quad 
    \sigma_{\mathrm{stat},ij}, 
    \quad (1 \leq i \leq n_{\mathrm{line}}, \; 1 \leq j \leq n_{\mathrm{cont}}).
\end{equation}
In short, for each annulus, we obtain a distribution of 100 independent $W_{\lambda}$(\lyaa) measurement, each reflecting a different choice of continuum and integration range.
We then take the mean of all $\mathrm{EW}_{ij}$ as our final equivalent width, and use their scatter, $\sigma_{\mathrm{sys}}$, as an estimate of the systematic uncertainty. The total error on the final measurement is given by
\begin{equation}
    \sigma_{\mathrm{EW}} \;=\; 
    \sqrt{\sigma_{\mathrm{stat}}^2 + \sigma_{\mathrm{sys}}^2},
\end{equation}
where $\sigma_{\mathrm{stat}}^2$ is the mean of the statistical variances over the $N$ measurements, and $\sigma_{\mathrm{sys}}^2$ is the variance in the $\mathrm{EW}_{ij}$ values themselves.

\begin{figure}
    \centering
    \includegraphics[height=5cm]{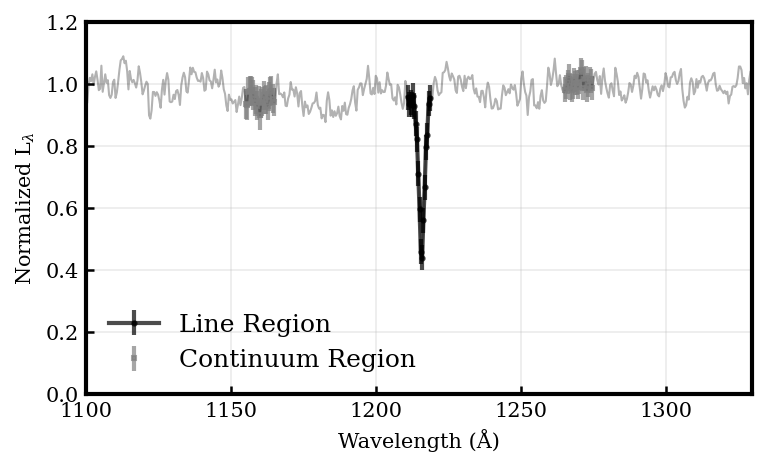}
    \caption{An illustration of the non-parametric equivalent width measurement for the 80–120 kpc annulus. The continuum is estimated from the gray regions blue-ward and red-ward of the absorption, and the integrated flux deficit shown in black defines W$_\lambda$(\lyaa). We do this for 100 unique combinations of definitions of line and continuum regions as discussed in Section \ref{sec:ew_measure}.}
    \label{fig:nonparam}
\end{figure}

Figure \ref{fig:nonparam} shows an example of this non-parametric measurement for the annulus spanning 80--120\,kpc. The continuum is defined as 1155\,\AA\ - 1165\,\AA\ on the blue side and 1265\,\AA\ - 1275\,\AA\ on the red side. The line region is defined as 1211\,\AA\ -1219\,\AA. The integrated flux deficit below this continuum provides an estimate of $W_{\lambda}$(\lyaa). We perform this procedure for all 100 unique combinations of line and continuum definitions, as discussed above. We also implement a Monte Carlo variant of the non-parametric measurement, using fixed definitions for the continuum and line integration regions. In this approach, we repeatedly perturb the flux values within their associated uncertainties, applying random Gaussian shifts with widths set by the local flux errors, to generate a distribution of equivalent widths. The resulting values match those from our primary method within 5\%. This Monte Carlo approach captures the statistical errors, and the approach described where we vary the continuum is intended to include both statistical and systematic.

The \lya absorption line is typically offset from the systemic redshift of the galaxy by $\sim 200$ $km/s$ \citep{Song_2014,Hashimoto_2015}. As a result of our reliance on the \lya emission redshifts, when shifting the individual spectra to the rest-frame based on the central \lya emission of each LAE, the absorption features may not be perfectly aligned prior to the final stacking. This misalignment introduces a statistical broadening of the stacked absorption profile. Since the absorption is not detectable in individual spectra, we are unable to correct for this velocity offset at this time. Finally, note that we apply a correction to our W$_\lambda$(\lyaa) measurement that originates from the fact that we do not and cannot continuum normalize our individual spectra prior to stacking. We discuss this in detail in Section \ref{subsubsec:bias}

\begin{figure*}
    \centering
    \includegraphics[height=12cm]{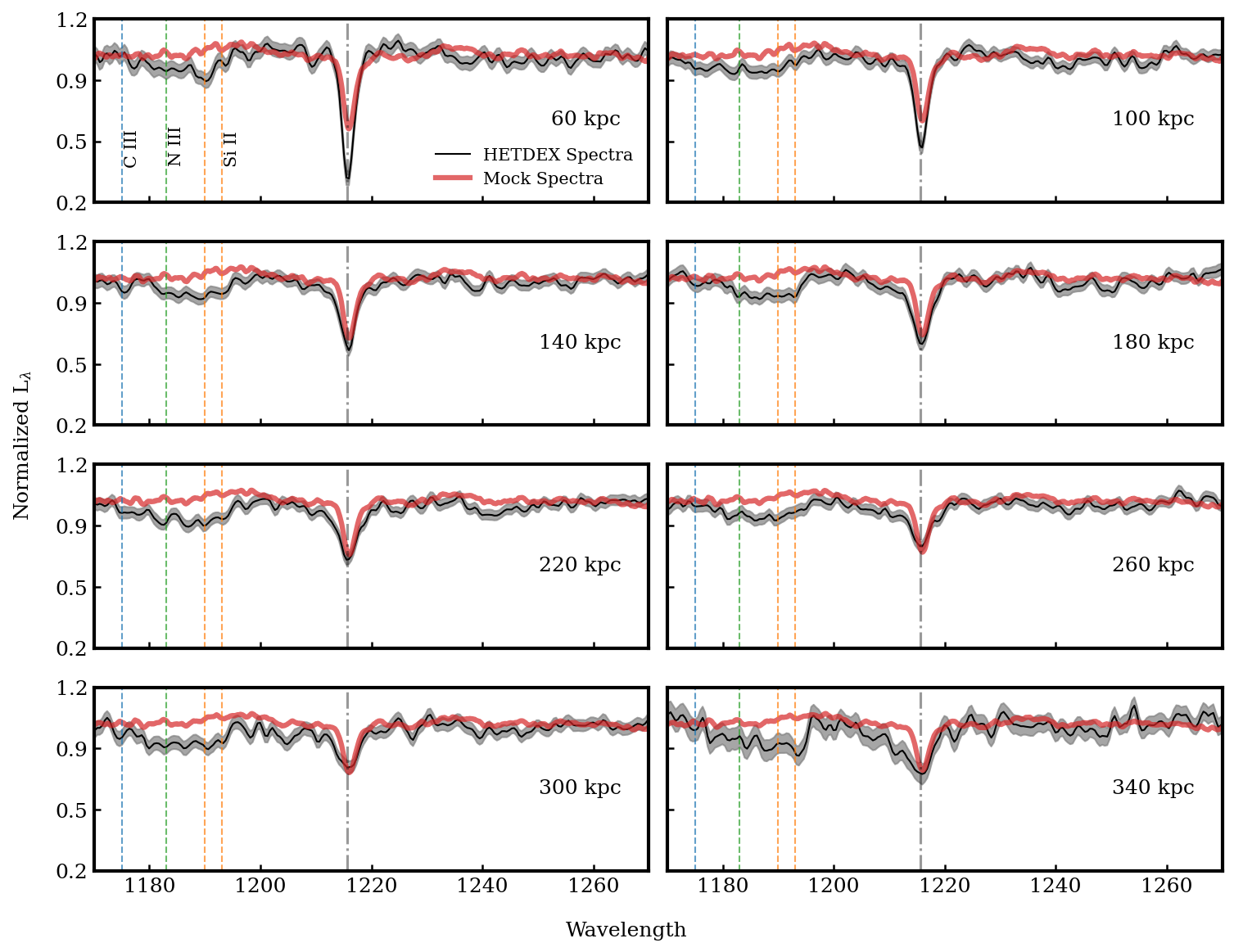}
    \caption{A comparison between stacked HETDEX spectra (black) and stacked \textsc{ASTRID} spectra (red) in eight 40\,kpc bins from $\sim$ 60\,kpc to 350\,kpc. The dot–dashed vertical line marks the \lya rest wavelength; the dashed lines indicate the expected positions of C \textsc{III} 1175 \,\AA, N \textsc{III} 1183 \,\AA, and the Si \textsc{II} 1190 \,\AA\, 1193 \,\AA\ doublet. A continuum suppression appears in the data between $\sim$ 1180 \,\AA\ - 1200 \,\AA\, yet is absent from the mock stacks. The most plausible explanation is extra absorption from metal lines in this interval, which are not included in our simulations. The ASTRID spectra are derived from forward modeling the HETDEX analysis steps, including merging 20 kpc annuli and the double biweight step. The simulation reproduces the overall depth and shape of the absorption profiles fairly well (See Section \ref{subsec:mocks})}.
    \label{fig:comparison}
\end{figure*}

\section{\textbf{Results}}
\label{sec:results}

Now that we have established the method for measuring $W_{\lambda}$(\lyaa) (except the correction for the bias for continuum normalization which we detail in Section \ref{subsubsec:bias}) , in this section, we compare our stacked HETDEX spectra to stacked spectra from hydrodynamical simulations. We then present the measured $W_{\lambda}$(\lyaa) profile of an average HETDEX LAE at z $\sim$ 2.5 and compare this profile with the one generated using the mock spectra. We then place our measurements in the context of other $W_{\lambda}$(\lyaa) studies to interpret our results within the broader framework of neutral hydrogen distribution around high-redshift star-forming galaxies.

\subsection{Comparison to Simulations}
\label{subsec:sim}

We compare our HETDEX-stacked spectra for each annulus to mock spectra generated from the \textsc{ASTRID} hydrodynamical simulations \citep{Bird_2021}. ASTRID is a cosmological smoothed-particle hydrodynamics (SPH) simulation, run with a modified version of GADGET-3, that evolves galaxies within a periodic box of 250 $h^{-1}$ cMpc, resolving them in halos with $M_{\mathrm{halo}} > 2 \times 10^9 M_{\odot}/h$ . The simulation incorporates essential physical processes such as multiphase star formation, metal cooling, and self-shielding of dense neutral gas, and adopts a uniform UV background to mimic reionization effects. Note that modeling the \lya emission line from the central region of the galaxies is not yet implemented, which is an aspect we aim to address in future work. The simulation reproduces the column density distribution function of high column density absorbers at z$\sim 2$-$3$ \citep{Bird2_023}.

From the full simulated sample, we select 840 galaxies that exhibit steep, wide absorption troughs with sharp edges in their central line-of-sight spectra. This selection is motivated by the findings of \cite{weiss_absorption_2024}, which show that the central spectra of HETDEX LAEs feature pronounced, box-like absorption profiles on both the blue and red sides of the \lya emission line. To identify comparable systems in the simulation, we extract the central line-of-sight spectra for all galaxies and perform a simple fit using a box-shaped absorption profile. The simulated galaxies that best match the observed absorption features form the final sample used for our comparison. The \lya absorption field is computed using \texttt{fake\_spectra} \footnote{\href{https://github.com/sbird/fake_spectra}{\texttt{https://github.com/sbird/fake\_spectra}}} package \citep{fake_spectra}, following procedures similar to those described in \cite{Qezlou_2023}.

\subsubsection{Mock Spectra}
\label{subsec:mocks}

For each galaxy we extract line‑of‑sight (LoS) spectra in concentric annuli, selecting 256 random lines of sight per annulus. Note that these extracted spectra are already in the rest frame—shifted according to the galaxies’ redshifts—and that the continuum in the simulated spectra is initially set at unity and drops by roughly 25\% because of mean \ion{H}{1} absorption in the IGM. The actual level of the continuum for the background light for each galaxy is an unknown parameter, and we do not try to model it since it is a secondary effect.

Each sight-line spectrum is converted from optical depth to flux and convolved with a Gaussian kernel that matches the VIRUS resolving power. To capture both LoS-to-LoS and galaxy-to-galaxy intrinsic variations in the background continuum, we multiply each spectrum by a factor drawn from a truncated $\mathcal{N}(1,0.5)$ distribution (positive only), corresponding to a $\sim50\%$ scatter motivated by the variations in the UV background light measurement reported by \citealt{Weiss_2025}. Following our HETDEX procedure (Sections~\ref{sec:stacking} and~\ref{sec:stacking_regions}), the 256 spectra in an annulus are biweight-averaged to create a single galaxy-averaged spectrum. These galaxy-averaged mock spectra are then biweight-stacked across all galaxies within each annulus.

After stacking, we define two continuum side‐band regions in the rest frame (one blueward and one redward of \lyaa) and take the median flux in each to linearly interpolate a continuum estimate. We then normalize each stacked mock spectrum by its continuum fit. Finally, as in our real data, we merge consecutive 20~kpc bins into 40~kpc bins to boost the signal-to-noise ratio in the outer regions. These additional steps—particularly merging 20~kpc bins and the two-stage biweight averaging—are primarily for forward-modeling the HETDEX procedure. Although omitting them yields nearly identical results, their inclusion is important for consistency with the observational analysis. We further note that, consistent with our treatment of the real data, we do not correct for the velocity offset of the absorbers and simply stack each line of sight as is.

Figure~\ref{fig:comparison} compares the stacks of mock spectra (red) and the HETDEX spectra (black) across eight 40~kpc bins spanning in $\sim 60-340$ kpc range. The dot–dashed vertical line marks the \lya rest wavelength; the dashed lines indicate the expected positions of C \textsc{III} 1175 \,\AA, N \textsc{III} 1183 \,\AA, and the Si \textsc{II} 1190 \,\AA\, 1193 \,\AA\ doublet. The data show a continuum suppression between $\sim$ 1180 \,\AA\ - 1200 \,\AA\ that is not reproduced by the mocks. This mismatch may arise, at least in part, from metal absorption associated with transitions in that region, which is not included in the mock spectra. Overall, the mock stacks reproduce both the depth and width of the observed absorption profile well, with minor discrepancies in the line shape, likely due to \lya velocity offsets and sky residual uncertainties.

\subsubsection{Correcting for Not Normalizing prior to Stacking}
\label{subsubsec:bias}

\begin{figure}
    \centering
    \includegraphics[height=5cm]{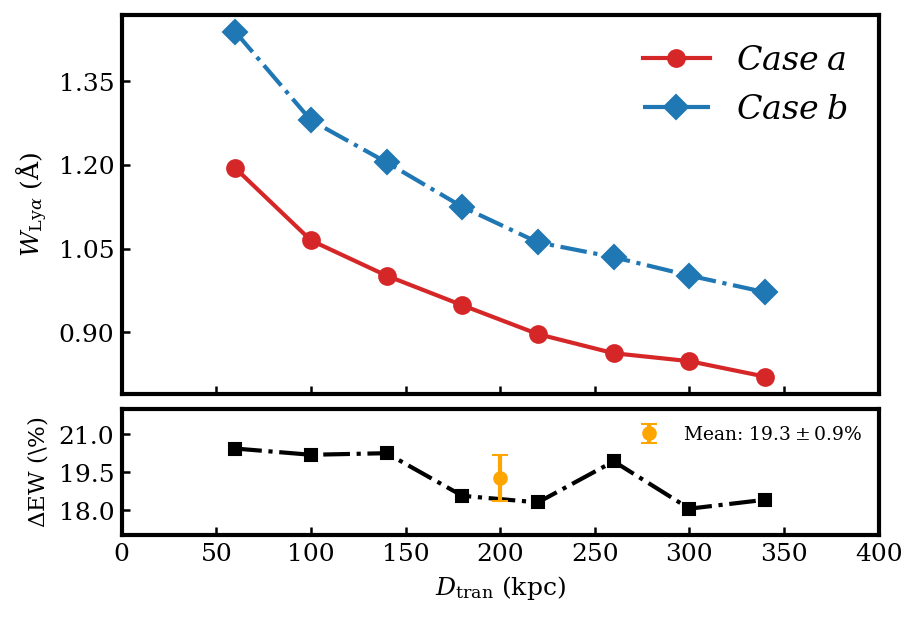}
    \caption{Impact of skipping per-spectrum continuum normalization. \textit{Top:} absolute \lya equivalent-width profiles for continuum-normalized stacks (Case~a, red) and HETDEX-style stacks without pre-normalization (Case~b, blue). \textit{Bottom:} fractional bias $(W_\lambda^{\rm b}-W_\lambda^{\rm a})/W_\lambda^{\rm a}$ in each 40 kpc bin. A $\sim 20\%$ uniform correction factor is applied to the observed HETDEX equivalent widths in all subsequent analysis.}
    \label{fig:bias}
\end{figure}

In standard absorption line and stacking analysis, every spectrum is first divided by its continuum, and only then is stacked for any subsequent $W_{\lambda}$ measurement.  For the majority of HETDEX LAEs the situation is different: aside from the \lya emission, the individual spectra contain no measurable continuum, and the noise in a single spectrum is too high to identify it. Consequently we stack first and only afterwards apply a continuum normalization to the stack spectrum. While unavoidable, this shortcut could bias the $W_{\lambda}$ since any galaxy-to-galaxy fluctuations in the intrinsic continuum level survive the stack.

Fortunately, the mock spectra allows us to estimate this bias. Using the same set of spectra ($\sim 215,000$ for each annulus) from the 840 galaxies detailed in Section \ref{subsec:sim}, we compare two cases. In case \textit{a} we follow the standard route: each simulated sight-line spectrum is divided by its own continuum before stacking, so the stacked result gives the ``true" $W_{\lambda}$. In case \textit{b}, as detailed in Section \ref{subsec:mocks} we imitate the analysis for the data by multiplying every sight-line spectrum by a random scale drawn from a truncated $\mathcal{N}(1,0.5)$ distribution(again capturing both LoS-to-LoS and galaxy-to-galaxy continuum scatter) then stacking without prior normalization. All subsequent steps (biweight averaging, pseudo-continuum fitting, $W_{\lambda}$ integration, 20kpc bin merging) are identical between the two cases, to make sure that any difference arises solely from skipping the per-spectrum continuum division.

Figure \ref{fig:bias} shows the $W_{\lambda}$(\lyaa) profile for the two cases. Across all eight 40 kpc bins the ``HETDEX-style" stacks (Case \textit{b}) yield $W_{\lambda}$(\lyaa) that are systematically larger by $19.3\%\pm0.9\%$. We therefore apply a uniform correction based on this $\sim 20\%$ bias to the measured HETDEX $W_{\lambda}$(\lyaa), so that the values reported in the paper represent the bias-corrected equivalent widths.

\subsubsection{Equivalent Width Profile}
\label{subsubsec:ew_profile_results}

\begin{figure}
    \centering
    \includegraphics[height=5cm]{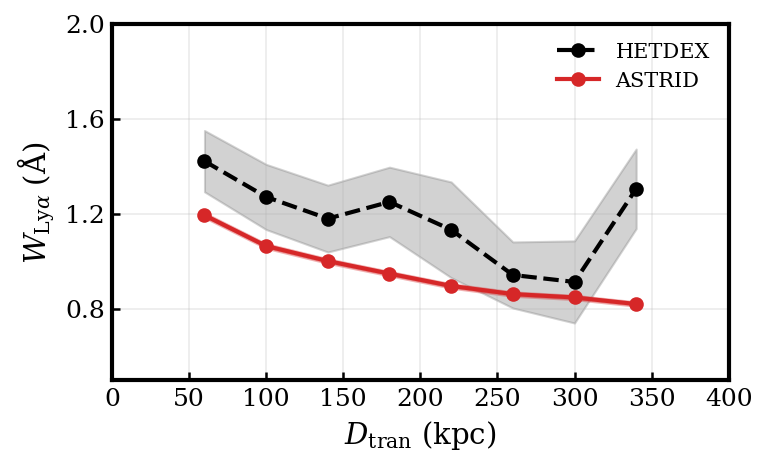}
    \caption{The bias corrected radial $W_{\lambda}$(\lyaa) profile of an average HETDEX LAE at $z \sim 2.5$ (black), along with the corresponding profile derived from the \textsc{ASTRID} mock spectra (red). For both datasets, the equivalent widths were measured non-parametrically (See Section \ref{sec:ew_measure}). Overall, the mock profile exhibits good qualitative agreement with the observed profile.}
    \label{fig:profile}
\end{figure}

Figure~\ref{fig:profile} shows the radial $W_{\lambda}$(\lyaa) profile of an average HETDEX LAE at $z \sim 2.5$ (black) after applying the $\sim 20 \%$ bias correction derived in Section~\ref{subsubsec:bias}, along with the corresponding profile derived from the \textsc{ASTRID} mock spectra (red), which is the red curve in Figure \ref{fig:bias}. The equivalent widths are measured non-parametrically (Section~\ref{sec:ew_measure}), using the same procedure for continuum estimation and integration. Overall, the mock profile exhibits good qualitative agreement with the observed profile, indicating that \textsc{ASTRID} effectively reproduces the diffuse \ion{H}{1} distribution around galaxies out to $\sim 350$~kpc. It is important to note that the galaxies from which the mock spectra are extracted are not necessarily LAEs. Our selection is based solely on matching the absorption features as described in Section \ref{subsec:sim}.

\subsection{Comparison to Literature}
\label{subsec:literature}

\begin{figure*}
    \centering
    \includegraphics[height=9.5cm]{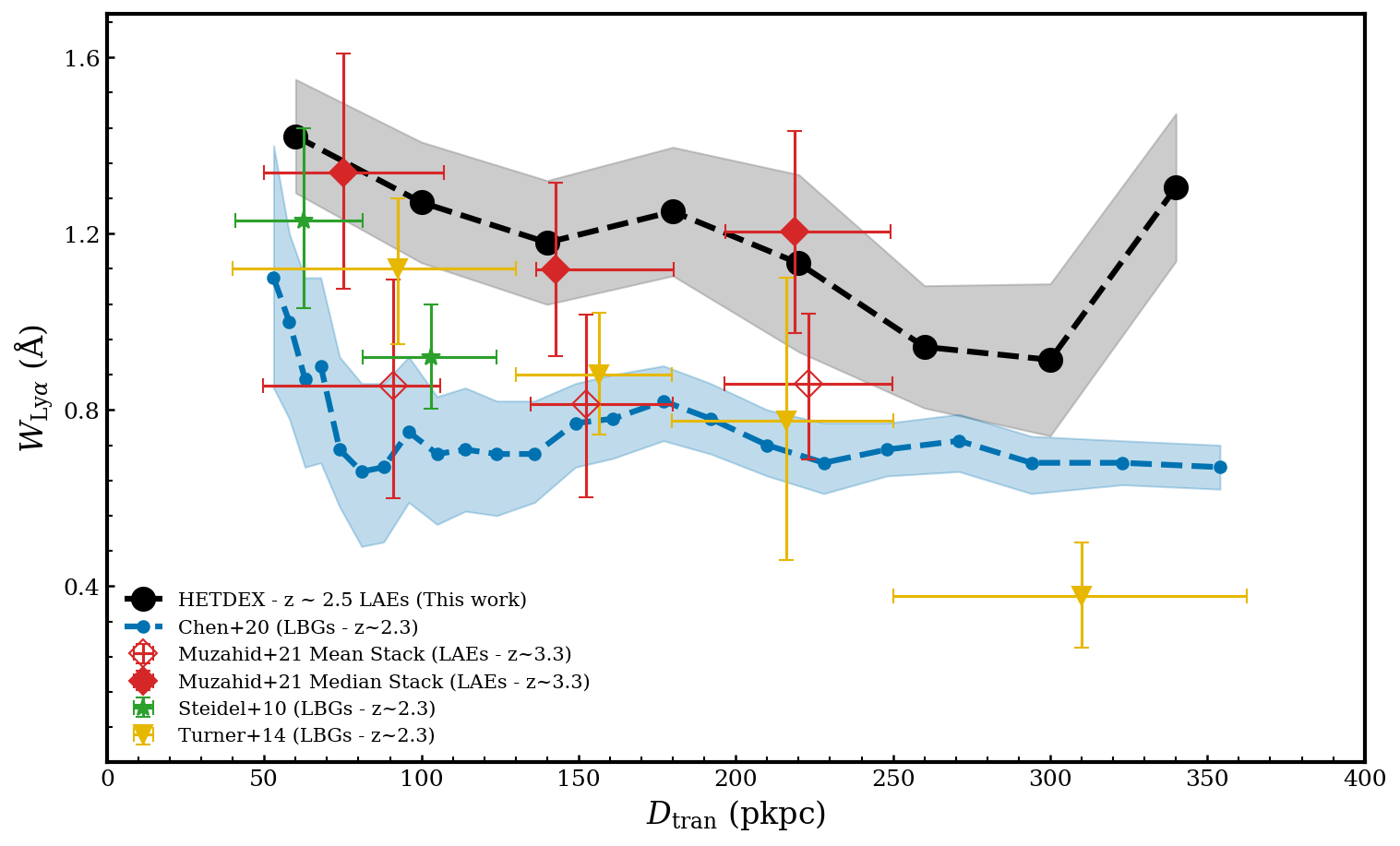}
    \caption{HETDEX radial $W_{\lambda}$(\lyaa) profile compared to the other studies. \textit{Black circles} with grey band: bias-corrected HETDEX LAEs (this work). \textit{Blue line} with shaded region: LBGs from \citet{Chen_2020}. \textit{Red hollow and filled diamonds}: LAE profiles from \citet{Muzahid_2021} measured on mean and median stacks, respectively. \textit{Green stars}: LBGs from \citet{Steidel_2010}. \textit{Yellow triangles}: LBGs from \citet{Turner_2014}. The HETDEX profile aligns with the median stacks of \citet{Muzahid_2021} LAE measurements but lies above the continuum-selected LBG curves.  A continuum-brightness subsample of HETDEX LAEs (not shown; see text) can move the profile toward the LBG curve, suggesting that selection effects, and possibly environmental factors, contribute to the difference.}
    \label{fig:study_compare}
\end{figure*}

Figure~\ref{fig:study_compare} places the bias–corrected HETDEX $W_{\lambda}$(\lyaa) profile (black circles) alongside a number of high–redshift studies. The blue curve shows the $z \sim 2.3$ Lyman Break Galaxy (LBG) profile reported by \citealt{Chen_2020}; green stars and yellow triangles are also individual $z \sim 2.3$ LBG data points from \citet{Steidel_2010} and \citealt{Turner_2014}, respectively; red hollow and filled diamonds are the mean and median-stack $z \sim 3.3$ LAE measurements of \citealt{Muzahid_2021}. After the continuum-bias correction (Section~\ref{subsubsec:bias}), the HETDEX LAE profile matches the \citealt{Muzahid_2021} z $\sim$ 3.3 \emph{median} stacks, yet lies relatively above the rest of the $W_{\lambda}$(\lyaa) profiles.

A trend within our own sample offers a promising clue on this difference. When we restrict the stack to LAEs that possess bright $r$-band continuum counterparts we obtain substantially lower $W_{\lambda}$(\lyaa) values; in the brightest bin of the sample the absorption even flips sign and becomes net \lya \emph{emission}.  Because LBG surveys select galaxies on the strength of their rest-UV continuum, they preferentially sample this bright-continuum regime, where our own stacks yield the weakest equivalent widths.  Selecting on continuum brightness therefore moves the LAE profile toward the LBG curve and could account for at least part of the observed offset. A deeper analysis will be required to determine how much of the difference is due to selection versus intrinsic CGM physics; we defer that investigation to a follow-up paper.

Intriguingly, \citealt{Muzahid_2021} show that $W_{\lambda}$(\lyaa) can be pushed even higher, reaching $\sim$ 1.77\,\AA\ in median stacks when the sight lines probe LAEs that belong to “groups'' residing in overdensities. Therefore, the location of an LAE within large-scale structure can be nearly as influential as its intrinsic properties in shaping the observed absorption. If the majority of HETDEX LAEs reside in density peaks, the additional neutral gas in those environments would naturally drive stronger absorption. In short, both local environment and selection effects appear to be key regulators of the observed $W_{\lambda}$(\lyaa) difference seen in Figure \ref{fig:study_compare}. We discuss this environmental effect more in Section \ref{subsec:model}.

It is worth noting that similar environmental trends have been seen in recent studies. \citealt{Momose_2021} reports that, despite LAEs typically residing in lower-mass halos compared to LBGs, their limited sample (19 LAEs) exhibits the highest cross-correlation function signal on small scales, which means denser surrounding \ion{H}{1} regions. Despite the fact that this result remains tentative due to the small number statistics, it hints at a another partial explanation for our findings; the elevated $W_{\lambda}$(\lyaa) observed in the full HETDEX sample may partly reflect an environmental bias, where LAEs are preferentially located in denser regions compared to LBGs.

\section{\textbf{Discussion}}
\label{sec:discussion}
Presenting a comprehensive physical model for this analysis requires modeling the central \lya emission from galaxy in a cosmological setting, which we reserve for a future paper. For now, we provide a simple phenomenological picture to explain the observed trends. Throughout this section, when we refer to S/N, we specifically mean the signal-to-noise ratio of the LAE's emission line.

\subsection{Physical Picture}
\label{subsec:model}

\begin{figure*}
    \centering
    \includegraphics[height=9.5cm]{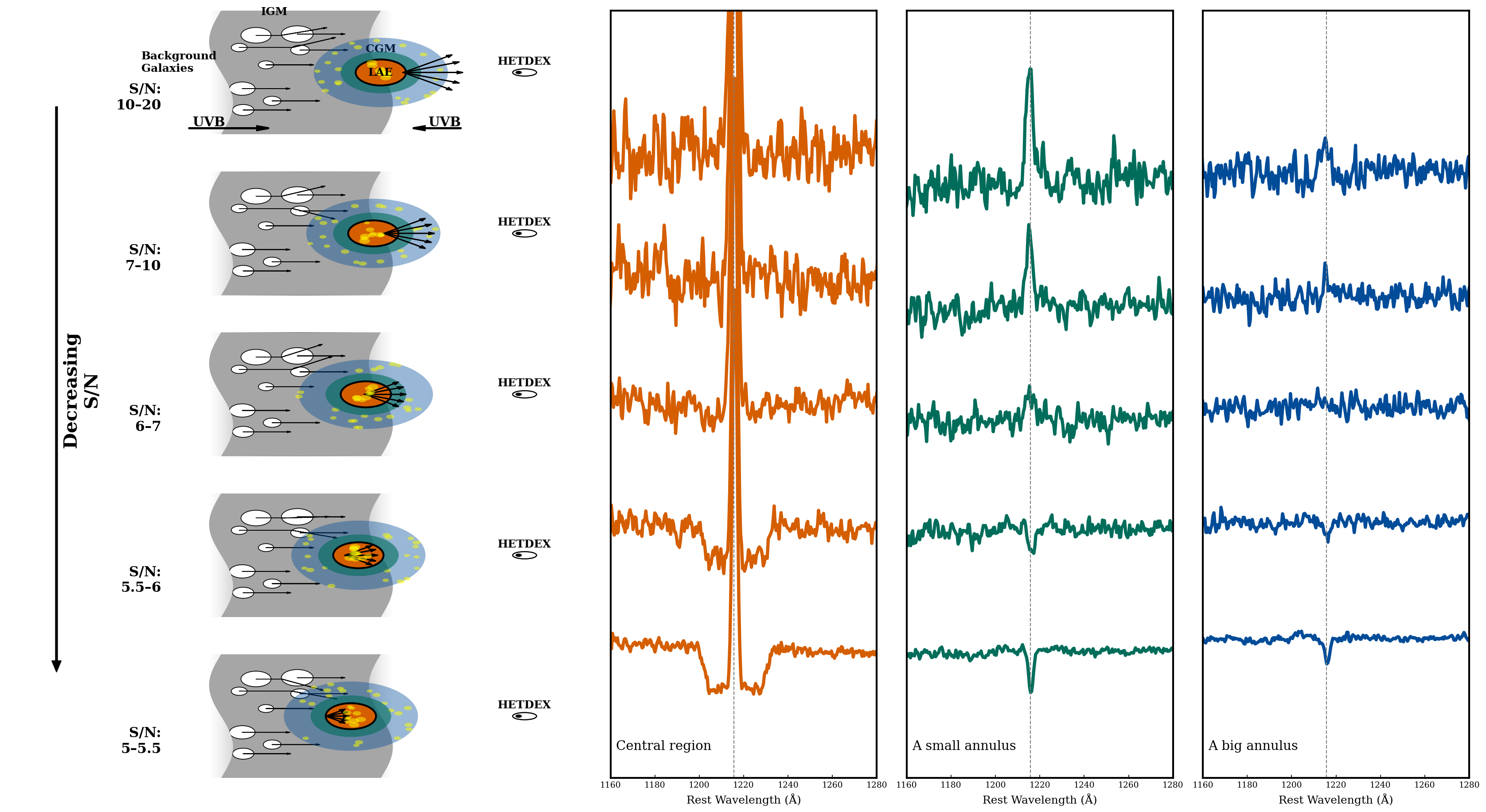}
    \caption{The stacked spectra as a function of distance from the LAE center and S/N. As demonstrated, absorption flips to emission in the stacks of annuli region (green and blue) and absorption troughs, discussed in \citealt{weiss_absorption_2024}, diminish in the stacks of aperture region (orange) as we increase S/N. Each line is separated by a vertical offset for display purposes. The depth of the troughs (orange) decreases in a manner similar to the weakening observed in \cite{weiss_absorption_2024} using luminosity function normalizations. The most left panel depicts a simple graphic (similar to \citealt{Weiss_2025}) of how a deeper embedding into dense environments can suppress the \lya emission and reduces the observed signal. We additionally illustrate gas and dust clumps (yellow circles) whose geometry determines the \lya escape fraction, further quenching emission. In this simple graphic, we show that as you embed the galaxy into a denser IGM and have more gas and dust surrounding the \lya emitting region within the galaxy, the \lya flux gets suppressed and the total amount of received signal decreases.}
    \label{fig:snr}
\end{figure*}

To begin understanding the physical picture, we stack spectra based on various properties of the LAEs. Figure \ref{fig:snr} presents stacked spectra as a function of distance from the LAE center and S/N. The left spectrum panel (orange) displays stacked spectra within the central emission line region, the central spectrum panel (green) covers a small annulus (40–60 kpc), and the right spectrum panel (blue) covers a larger annulus (80–100 kpc). For clarity, each spectrum is assigned an arbitrary vertical offset for display purposes.  Moving from bottom to top, S/N is increasing in all panels. As the stacks in the left panel of Figure \ref{fig:snr} show, the \lya absorption lessens with increasing S/N in the central region of the LAE. In annuli around the LAE (middle and right panels of Figure \ref{fig:snr}), the absorption lessens with increasing S/N, but at sufficiently high S/N, \lya \textit{emission}  is observed instead. The leftmost panel shows a simple graphic which could explain the trends we see here, and we expand upon that in this section.

Note that we use S/N as a parameter since it appears to drive the observed differences. We acknowledge that the apparent luminosity of the LAEs is a more direct physical parameter for this analysis. Therefore, we perform binning and stacking of the LAEs based on their luminosity as well. We indeed find that the absorption-to-emission behavior around LAEs also occurs with increasing luminosity. However, when we bin by S/N within each luminosity bin, we still observe this absorption-to-emission trend. This observation suggests that S/N, rather than luminosity, is the main determinant.

Extended \lya emission and absorption around star-forming galaxies have been observed in spectral stacking analyses \citep{niemeyer_surface_2022,lujan_niemeyer_lyensuremathalpha_2022}, deep field long exposure observations \citep{leclercq_muse_2020,Kikuchihara_2022}, and quasar sightline studies \citep{Mukae_2020}. While in this paper we detect \lya absorption features around HETDEX LAEs, \citeauthor{niemeyer_surface_2022} instead report emission around HETDEX LAEs. The difference is due to the chosen S/N thresholds. \citeauthor{niemeyer_surface_2022} use a minimum S/N of $6.5$, whereas we use a minimum S/N of $5$. Their result is consistent with the trend presented in Figure \ref{fig:snr}, since the absorption is only evident at $5\lesssim\mathrm{S/N}\lesssim6$. We note that the majority of HETDEX LAEs fall within this lower S/N bin.

According to \citealt{weiss_absorption_2024} \& \citealt{Weiss_2025}, a possible geometry that explains the observed \lya absorption troughs in the central region of HETDEX LAEs is as follows: when an LAE is partially embedded in an \ion{H}{1} screen (mimicking the foreground edge of an overdensity or filament), flux from background sources gets absorbed by the \ion{H}{1}. Observations of the LAE (which include this \ion{H}{1} halo) result in a combined profile of central \lya emission sitting in an absorption well. The degree to which the LAE is embedded in the \ion{H}{1} inversely correlates with the observed signal; meaning that for a fixed luminosity, the deeper an LAE is embedded, the lower the observed signal. This picture suggests a correlation between the S/N ratio and the depth that an LAE is embedded within these \ion{H}{1} regions. Based on Figure \ref{fig:snr}, strongest absorption in the annuli is seen around the LAEs that exhibit deep troughs in their central region. Another factor contributing to the observed absorption in LAEs with lower S/N is that LAEs in overdense regions experience higher levels of photon flux and greater variations in these flux levels due to the denser environment. This increased photon flux and fluctuation results in elevated noise levels in the observations. Using luminosity function normalizations as a proxy for density enhancements, as in \cite{weiss_absorption_2024}, we observe a higher mean noise levels for LAEs located in overdense fields. 

We also note that \lya emission can be quenched internally by dust and \ion{H}{1} within the galaxy’s ISM: when star-forming knots lie deeper within dusty clumps, the intrinsic \lya escape fraction is reduced, diminishing both central and halo emission. As a result, we primarily attribute the observed absorption in lower S/N LAEs to these factors: (1) the LAE being more embedded in and/or residing in denser \ion{H}{1}, leading to a suppression of the \lya emission and hence lower observed signal (see Figure \ref{fig:snr}); (2) the LAE’s location in a locally overdense region, leading to an elevated noise level; and (3) \lya quenching by internal ISM dust, which lowers the intrinsic escape fraction and further suppresses the \lya emission. We illustrate all of these effects in the leftmost graphic of Figure \ref{fig:snr}. In practice, we think S/N acts as a proxy for a combination of physical drivers which are IGM \ion{H}{1} absorption, environmental noise, and internal dust quenching that together regulate the net \lya signal.

\subsection{Reliability of the Absorption Feature}
\label{subsec:reliability_absorption}

To verify that the observed absorption signals are real and not artifacts of instrumental effects, false positives, or sky subtraction issues, we perform several consistency checks. First, when we stack spectra from the fibers surrounding low redshift [O II]-emitting galaxies within the HETDEX dataset instead of LAEs, we do not see any absorption features at \OII wavelength, suggesting that the absorption is specific to LAEs and the \lya wavelength. This finding is in agreement with \citealt{weiss_absorption_2024} who do not detect absorption troughs in the central region of the \OII-emitting galaxies. Second, the troughs and absorption are present in the original spectra without sky subtraction for both the central and surrounding region of the LAEs, indicating that it is not a signal introduced during sky subtraction \citep{Weiss_2025}. 

\begin{figure}
    \centering
    \includegraphics[height=4cm]{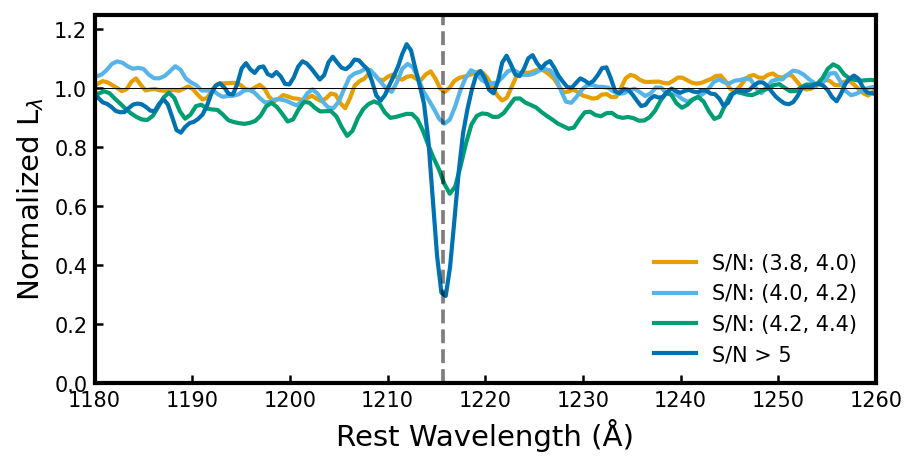}
    \caption{Stacked, continuum-normalized spectra of LAEs at a small annulus of 40-60 kpc divided by S/N bins in the range of 3.8-5. The vertical dashed line marks the rest-frame \lya wavelength. The stacks are smoothed with a 1-pixel Gaussian kernel. The absorption signal is prominent for S/N~$>5$ but weakens toward lower S/N, demonstrating that the feature is not produced by an increasing rate of false positives at low S/N.}
    \label{fig:low_sn}
\end{figure}

A further concern is that an increased fraction of false-positive detections at low S/N might create a fake net absorption in the stacked spectra.  If this were the case, the strength of the absorption should grow—not fade—as we move to progressively lower S/N detections. Figure \ref{fig:low_sn} shows the opposite: the strength of the \lya\ absorption diminishes and is no longer significant in the very low S/N stacks. The trend confirms that false positives dilute, rather than create, the absorption signal. It is only after ruling out these possible instrumental and data reduction issues that we proceed with our analyses of the absorption signal.

\subsection{Anisotropy \& Observational Biases}
\label{subsec:antisotropy}

It is expected that the statistical continuum source that we observe getting absorbed at \lya would be isotropic. Due to HETDEX's observational bias towards LAEs that are not obstructed by gas or dust (see Figure 8 of \citealt{weiss_absorption_2024}), we observe a local anisotropy in both the \ion{H}{1} distribution and the \lya\ photon flux along the line of sight, which makes the detection of absorption from these LAEs possible. The underlying physical model of these observations is not fully understood and is the subject of ongoing and future works. However, we have demonstrated that the detected absorption signal is adequate and robust enough to motivate further modeling and analysis.

Additionally, in interpreting the results, it is crucial to consider the limitations imposed by our reliance on the redshifts from the \lya\ emission line. As detailed in Section \ref{sec:ew_measure}, when we shift the spectra to the rest frame (which is determined via \lya\ emission from the galaxy center), the absorption features do not align perfectly prior to stacking. This misalignment leads to a statistical broadening of the final stacked absorption profile. As of this work, we cannot correct for this velocity offset since the absorption is too faint in individual spectra. It is possible that this statistical broadening is another reason that our measured $W_{\lambda}$(\lyaa) values are relatively higher than those reported by some of the studies mentioned in Section \ref{subsec:literature}. The broadening may produce a profile that is overall shallower yet wider, leading to a higher integrated equivalent width.

\subsection{Future Directions}
\label{subsec:future}

An important next step is to study the variation of the $W_{\lambda}$(\lyaa) profile with local density, using luminosity function normalizations as a probe for environmental differences, similar to those used in \cite{weiss_absorption_2024} (see Figure 6 of the paper). Despite the current large number of LAEs, the available sample is not yet sufficient for this analysis; the HDR5 release, which will include all HETDEX data, is expected to make this effort possible. 

That being said, the detection of \lya absorption using this new method provides evidence for substantial \ion{H}{1} gas in the circumgalactic medium around LAEs at $z \sim 2.5$. We aim to quantify the extent and density of the \ion{H}{1} gas around different populations of LAEs as a function of redshift, \lya luminosity, and local density. In particular, it is crucial to fully understand why the absorption appears predominantly in LAEs with $5 \lesssim S/N \, \lesssim 6$ -- corresponding to the lower end of the S/N range of our high-confidence LAEs and encompassing the majority of LAEs in HETDEX.

\section{\textbf{Summary}}
\label{sec:summary}
The extensive number of spectra from LAEs and their surroundings from HETDEX enabled us to investigate the distribution of \ion{H}{1} in the CGM of LAEs through a novel application of spectral stacking that leverages the integrated light of background galaxies along the line of sight. In this study, we analyze \lya absorption around $\sim$88,000 LAEs using data from HETDEX Data Release 4.0.0. We provide an empirical  $W_{\lambda}$(\lyaa) profile around an average LAE at $z \sim 2.5$, detecting \lya absorption in the range of $40 \lesssim D_{tran} \, (\mathrm{pkpc}) \lesssim 350$.  

Comparisons with \textsc{ASTRID} hydrodynamical simulations (Section \ref{subsec:sim}) show a fairly consistent $W_{\lambda}$(\lyaa) profile qualitatively to the one we observe around HETDEX LAEs. Comparison with the literature (Section \ref{subsec:literature}) reveals broadly a similar profile, though systematic differences are apparent that could possibly be linked to the intrinsic differences in properties of LAEs and LBGs and/or the selection and local density (environmental) effects. These findings underscore the potential of this new method to constrain the \ion{H}{1} distribution around high-redshift star-forming galaxies. We argue that the observed S/N trend (Figure \ref{fig:snr}) indirectly reflects underlying physical processes: LAEs situated in overdense regions of the large-scale structure are subject to stronger \ion{H}{1} absorption. Also LAEs whose their \lya emitting region lies deeper into dusty ISM clumps will have lower intrinsic \lya escape fractions. Both of these effects lead to suppression and quenching of \lya emission in the central region and the halo of the LAE. This points to a significant environmental and geometrical influence on the observed trend, as shown in the left panel of Figure \ref{fig:snr}. We plan to expand the analysis by incorporating central \lya emission from the galaxy into the simulations, refining the model to reflect the complex interplay between star formation, gas and dust content, and environment of these systems.

\acknowledgments
We thank the referee for their insightful feedback and constructive suggestions, which improved the clarity and quality of the paper. 

HETDEX is led by the University of Texas at Austin McDonald Observatory and Department of Astronomy with participation from the Ludwig-Maximilians-Universit\"at M\"unchen, Max-Planck-Institut f\"ur Extraterrestrische Physik (MPE), Leibniz-Institut f\"ur Astrophysik Potsdam (AIP), Texas A\&M University, The Pennsylvania State University, Institut f\"ur Astrophysik G\"ottingen, The University of Oxford, Max-Planck-Institut f\"ur Astrophysik (MPA), The University of Tokyo, and Missouri University of Science and Technology. In addition to Institutional support, HETDEX is funded by the National Science Foundation (grant AST-0926815), the State of Texas, the US Air Force (AFRL FA9451-04-2-0355), and generous support from private individuals and foundations.

Observations were obtained with the Hobby-Eberly Telescope (HET), which is a joint project of the University of Texas at Austin, the Pennsylvania State University, Ludwig-Maximilians-Universit\"at M\"unchen, and Georg-August-Universit\"at G\"ottingen. The HET is named in honor of its principal benefactors, William P. Hobby and Robert E. Eberly.

VIRUS is a joint project of the University of Texas at Austin, Leibniz-Institut f\"ur Astrophysik Potsdam (AIP), Texas A\&M University (TAMU), Max-Planck-Institut f\"ur Extraterrestrische Physik (MPE), Ludwig-Maximilians-Universit\"at Muenchen, Pennsylvania State University, Institut fur Astrophysik G\"ottingen, University of Oxford, and the Max-Planck-Institut f\"ur Astrophysik (MPA). In addition to Institutional support, VIRUS was partially funded by the National Science Foundation, the State of Texas, and generous support from private individuals and foundations.

The authors acknowledge the Texas Advanced Computing Center (TACC) at The University of Texas at Austin for providing high performance computing, visualization, and storage resources that have contributed to the research results reported within this paper. URL: http://www.tacc.utexas.edu

The Institute for Gravitation and the Cosmos is supported by the Eberly College of Science and the Office of the Senior Vice President for Research at the Pennsylvania State University.

KG acknowledges support from NSF-2008793.  EG acknowledges support from NSF grant AST-2206222. ASL acknowledges support from Swiss National Science Foundation. 
SS acknowledges the support for this work from NSF-2219212. SS is supported in part
by World Premier International Research Center Initiative (WPI Initiative), MEXT, Japan.

\facility{HET}

\software{Astropy \citep{astropy:2018}, NumPy \citep{numpy}, SciPy \citep{SciPy}, Matplotlib \citep{matplotlib}, EliXer \citep{davis_hetdex_2023}}


\clearpage

\bibliography{HI_density}


\end{document}